\documentclass[a4paper,10pt]{article}

\title{Flavor Mixing, Neutrino Masses\\and\\Neutrino Oscillations}
\author{H. Fritzsch\\University of Munich, Physics Department\\Munich, Germany}

\begin{document}

\maketitle

\begin{abstract}
We study a model for the mass matrices of the leptons. We are able to relate 
the mass eigenvalues of the charged leptons and of the neutrinos to the mixing 
angles and can predict the masses of the neutrinos. We find a normal hierarchy 
- the masses are 0.004 eV, 0.01 eV and 0.05 eV. The atmospheric mixing angle 
is given by the mass ratios of the charged leptons and the neutrinos. We find 
38 degrees, consistent with the experiments. The mixing element, connecting 
the first neutrino with the electron, is found to be 0.05. 
\end{abstract}
\bigskip
\bigskip
We observe three lepton-quark families in nature. The first family consists of 
the electron, its neutrino and the u- and d-quarks. The members of the second 
family are the muon, its neutrino and the c- and s-quarks. The third family 
consists of the tauon, its neutrino and the t- and b-quarks. It is unknown, 
whether there is a connection between the number of families and the number of 
colors in quantum chromodynamics.

In the Standard Model 28 fundamental constants have to be introduced. Their 
values cannot be calculated - they have to be measured in the experiments. 
Theses constants are:
\begin{itemize}
\item the constant of gravity G,
\item the fine structure constant,
\item the coupling constant of the weak interactions,
\item the coupling constant of the strong interactions or the scale of QCD,
\item the mass of the W-boson,
\item the mass of the ``Higgs''-boson,
\item the masses of the three charged leptons,
\item the masses of the three neutrinos,
\item the masses of the six quarks,
\item the four parameters, describing the flavour mixing of the quarks and
\item the six parameters, describing the flavour mixing of the leptons.
\end{itemize}
For the masses of the quarks we assume the following values:
\begin{center}
$u: \;5.3 \;\;\;\;\,MeV,$\\
$d: \;7.8 \;\;\;\;\,MeV,$\\
$s: \;146 \;\;\;\,MeV,$\\
$c: \;1050 \;\;\, MeV,$\\
$b: \;4 600 \;\;\;\, MeV,$\\
$t: \;174000\; MeV.$
\end{center}

The quark masses are scale dependent. The values given above, except the one 
for the t-mass, are normalized at an energy scale of 1 GeV. The experimental 
errors are not given, but are at least 10 \%, except for the t-mass, which is 
known to about 1 \%.

The quarks of the same charge do mix. If the u-quark interacts with a W-boson, 
a mixture of d, s and b appears. These mixtures are described by the CKM matrix 
(ref.1). I prefer a description, which we introduced some time ago (ref. 2):
\begin{equation}
V=\left[\begin{array}{ccc}c_u&s_u&0\\-s_{{u}}&c_{{u}}&0\\0&0&1\end{array}\right]
\left[\begin{array}{ccc}e^{-i\phi}&0&0\\0&c&s\\0&-s&c\end{array}\right]\left
[\begin{array}{ccc}c_{{d}}&-s_{{d}}&0\\s_{{d}}&c_{{d}}&0\\0&0&1\end{array}
\right]
\end{equation}
In the case of three families there are three mixing angles and one phase 
parameter. The latter describes the CP-violation. The angles with the index u 
or d describe the mixing in the u-c sector or the d-s sector, the angle with 
no index describes the mixing between the (t,c)-system and the (b,s)-system. 

I proposed years ago a simple texture 0 mass matrix for the quarks, given by 
the following matrix, valid both for the (u,c,t)-system and the  (d,s,b)-system 
(ref.3):

\begin{equation}
M=\left[ \begin{array}{ccc}0&A&0\\A^{{*}}&C&B\\0&B^{{*}}&D\end{array}\right]
\end{equation}
Such mass matrices are obtained, if in the electroweak theory special 
symmetries are present, either discrete reflection symmetries or continuous 
symmetries (ref.3). After diagonalization one finds the following relations 
for the mixing angles (ref. 4):
\begin{equation}
\theta_d=\tan^{-1}\sqrt{m_d/m_s}\;\;\;\;\;,\;\;\;\;\;
\theta_u=\tan^{-1}\sqrt{m_u/m_c}
\end{equation}
Taking as input the quark masses given above, we obtain for the mixing angle in 
the d-s- sector 13 degrees, and in the u-c-sector 4 degrees. The first angle is 
identical to the measured Cabibbo angle. These angles agree very well with the 
experimental data on the flavor mixing, if the phase angle is assumed to be 
close to 90 degrees. 

Analogously we can describe the flavour mixing in the lepton sector, which can 
be studied in the neutrino oscillations. The experiments, carried out in Japan 
(Kamiokande detector, ref. 5) and in Canada (SNO detector, ref. 6), give the 
following results for the mass-squared differences of the three neutrino mass 
eigenstates:
\begin{equation}
\Delta m_{21}^2\simeq 8\times10^{-5} eV^2\;\;\;\;\;,\;\;\;\;\;\Delta m_{32}^2
\simeq 2.5\times10^{-3} eV^2
\end{equation}
In neutrino oscillations only mass squared differences can be measured. No
information can be obtained for the absolute magnitude of the neutrino masses.
There is the possibility that the neutrino masses are nearly degenerate, e.g.
masses like $m_1 = 0.94  eV$, $m_2 =   0.95 eV$, $m_3 = 1   eV$ 
are possible. If one neutrino remains massless, one would have the masses 
$m_1 = 0eV$,$m_2=0.009eV$, $m_3 = 0.05  eV$. In the latter case a hierarchy 
of the masses is present, but this hierarchy is much weaker than the mass 
hierarchy for  the charged leptons. Below we shall calculate the neutrino 
masses.

Neutrino oscillations arise, since the neutrinos, produced by the weak 
interactions, are not mass eigenstates, but mixtures of mass eigenstates. 
Like for the quarks on has a 3x3 mixing matrix, which can be written as a 
product of three simple matrices, and a phase matrix, which is present only, 
if neutrinos are Majorana particles:
\begin{center}
${V=U\cdot P}$
\end{center}
\bigskip
\begin{equation}
U=\left[ \begin{array}{ccc}c_l&s_l&0\\-s_l&c_l&0\\0&0&1\end{array}\right]\left
[\begin{array}{ccc}e^{-i\phi}&0&0\\0&c&s\\0&-s&c\end{array}\right]\left
[\begin{array}{ccc}c_\nu&-s_\nu&0\\s_\nu&c_\nu&0\\0&0&1\end{array}\right]
\end{equation}
\begin{center}
$P=\left[\begin{array}{ccc}e^{i\rho}&0&0\\0&e^{i\sigma}&0\\0&0&1\end{array}
\right] \nonumber$
\end{center}
Here $s_\nu$ stands for $\sin {\theta_\nu}$ ($\theta_\nu$ : solar mixing 
angle), s  stands for $\sin {\theta}$ ($\theta$: atmospheric mixing  angle), 
and $ s_l $  stands for $\sin {\theta_l} $ ($\theta_l$ : reactor  mixing  
angle). The latter has  not been measured. The experimental results for the 
solar and the atmospheric mixing angles are (ref.5,6):
\begin{equation}
30^o \leq \theta_\nu \leq 39^o\;\;\;\;\;,\;\;\;\;\; 37^o\leq \theta\leq 53^o
\end{equation}
We assume for the lepton mass matrices the same texture 0 pattern as for the 
quarks (ref.7). Thus we obtain the same relations between the mixing angles 
and the mass eigenvalues: 

\begin{equation}
\tan {\theta_l}=\sqrt{m_e/m_\mu}\simeq0.07\;\;\;\;,\;\;\;\tan{\theta_\nu}=
\sqrt{m_1/m_2}
\end{equation}
Since the solar angle has been measured (we take 33 degrees), we find for the 
mass ratio of the first two neutrinos:
\begin{equation}
m_{1}/m_{2} \approx 0.42
\end{equation}
The oscillation experiments provide us with the mass squared differences. The 
new relation (7) allows us to determine the neutrino masses. We find (in eV):
\begin{center}
$m_1\approx0.004\;\;\;\;,\;\;\;\;m_2\approx0.01\;\;\;\;,\;\;\;\;m_3\approx0.05$
\end{center}
These neutrino masses are very small. We observe that the masses show a rather 
weak hierarchy, but the mass spectrum is not inverted. The first neutrino has 
the smallest mass.

If the neutrino masses are Majorana masses, one expects a neutrinoless double 
beta decay. The present limit on the Majorana mass is about 0.23 eV (ref. 8). 
This limit needs to be improved by about a factor 5. At this level the decay 
should be seen. 

The atmospheric mixing angle is consistent with 45 degrees. The parameter C in 
eq. (2) might be zero, as originally assumed (see ref. 3). But the high mass of 
the t-quark does not allow this possibility for the quarks. It might work for 
the leptons. In this case the atmospheric mixing angle is related to the two 
angles, which are given by the corresponding mass ratios:
\begin{equation}
\tan {\theta_1}=\sqrt{m_\mu/m_\tau}\;\;\;\;,\;\;\;\tan{\theta_2}=\sqrt{m_2/m_3}
\end{equation}
\begin{center}
$\theta_1\approx 14^o\;\;\;\;,\;\;\;\;\theta_2\approx 24^o$
\end{center}

The atmospheric mixing angle is given by the absolute value of the sum of the 
two angles, including a relative phase between the two terms. In order to get 
the direct sum, this phase must be 180 degrees. In this case we have for the 
atmospheric mixing angle:
\begin{equation}
\theta = \theta_1 + \theta_2 \simeq 38^o
\end{equation}
We can not obtain a maximal mixing (45 degrees), but our result is consistent 
with the experiment.

We can predict the matrix element  $V_{3e}$ of the mixing matrix V:
\begin{equation}
V_{3e}=\sin {\theta} \sin {\theta_l}\approx0.707\sqrt{m_e/m_\mu}\approx 0.05
\end{equation}
A matrix element of this magnitude could be observed in the upcoming reactor 
neutrino experiments.

\end{document}